# Electrical Signatures of Corrosion and Solder Bond Failure in c-Si Solar Cells and Modules

Reza Asadpour, *Student Member, IEEE,* Xingshu Sun, and Muhammad. Ashraful Alam, *Fellow, IEEE*
School of Electrical and Computer Engineering, Purdue University, West Lafayette, IN 47907

*Abstract*—Moisture- and temperature-activated corrosion of metal fingers, mechanical stress induced delamination, and failure of solder bonds rank among the leading failure mechanisms of solar modules. The physics of moisture ingress, diffusion and reaction have been explored in detail, but the electrical implications of corrosion and delamination on specific front-surface grid geometry is not fully understood. In this paper, we show that the module efficiency loss due to corrosion, delamination, and solder bond failure (CDS) involves a complex interplay of voltage/current redistribution, reflected as a loss in photocurrent as well as decrease/increase in shunt/series resistances. Our work will redefine the interpretation of experimental J-V characteristics features due to degradation mechanisms, integrate a variety of scattered and counter-intuitive experimental results within a common theoretical framework, and inform CDS-resistant grid design for solar modules.

## I. INTRODUCTION

The levelized cost of electricity (LCOE) of solar energy can be reduced by lowering the cost of manufacture, increasing the efficiency of the cells, and enhancing the reliability of the modules. Thus, the ability to predict the lifetime and improve the reliability of modules plays a pivotal role in commercial PV systems. Among various reliability issues (e.g., yellowing, PID, partial shading, etc.), Jordan *et al*. identified Internal Circuitry (IC) discoloration caused by corrosion as the second most significant degradation mode in the systems installed in the last 10 years [1]. Indeed, corrosion, delamination, solder bond failure (CDS) have always played a critical role in defining module lifetime [2], [3].

PV degradation can be monitored and predicted in one of the two ways: pre-installation accelerated qualification tests and post-installation off-line field tests. In accelerated qualification tests, well-controlled environmental stressors (e.g., humidity, temperature) attempt to isolate/accelerate the specific degradation pathway (e.g., UV test for yellowing) [4]–[11]. In practice, a given combination of stress conditions may in fact accelerate more than one degradation modes. Misattribution of multiple degradation modes to a single presumed degradation mechanism makes predictive modeling difficult. Specifically, correlated degradation makes interpretation of J-V characteristics challenging. For example, depending on stress condition used, a degradation mode (e.g. corrosion) may appear as a decreased shunt-resistance ($R_{sh} \sim (dJ/dV)^{-1}_{V=0}$), loss of short-circuit current ($J_{SC} \equiv J(V = 0)$), and/or increase in series resistance ($R_s \sim (dJ/dV)^{-1}_{V=V_{oc}}$), as has been reported in many experiments [4], [6], [11]. Any approach that considers $R_s$-increase as the sole signature of corrosion will miss important signs of early degradation related to finger corrosion, for example. The situation is even more complicated for off-line testing of fielded modules. Here the environmental stress factors are uncontrolled, and therefore multiple degradation modes occur simultaneously. The concurrent degradation mechanisms make it difficult to isolate degradation modes [1], [3] based on J-V analysis by a traditional five-parameter model.

In this paper, we will establish a new physics-based approach to interpret J-V signatures for degradations involving corrosion, delamination, and solder-bond failure (CDS). These J-V signatures will simplify the interpretation of accelerated tests as well as off-line field data. We wish to emphasize that significant amount of physical modeling and material characterization work have already been done to establish the kinetics of moisture diffusion, the physics of Na ion transport, and the reaction products formed near the contact during CDS degradation [12], [13]. However, the implications of these degradations in terms of module J-V characteristics are not clear. Therefore, in this paper, we wish to explore the physics of CDS and how they influence the electrical performance of the cells and modules. Then we will recommend a set of rules to differentiate these mechanisms and their features from other degradation modes. In Sec. II, we explain the simulation framework to study CDS failure effects. In Sec. III, we explain the corrosion geometry, solder bond failure, and their effects on

J-V curves of the affected cell. Then we use the cell results to simulate J-V of modules to demonstrate the effects in module level in Sec. IV. We discuss the results in Sec. V and conclude in Sec. VI.

## II. Theory and Modeling Framework

Solar cell consists of a substrate that absorbs sunlight and generates electron-hole pairs. Photo-generated carriers are transported away from the cell by metal contacts. In a c-Si cell, the front metal contacts are arranged in a hierarchical grid pattern to balance shading of the incident light vs. power lost to "series" resistance during charge collection. Typically, the H-shaped grid consists of (~100 μm) thin fingers (red vertical lines, Fig. 1a) carrying the current from semiconductor towards a thicker (~1mm) busbar (thick horizontal blue lines, Fig. 1a). The busbars themselves are contacted with ribbons (thinner horizontal yellow lines, Fig. 1a) via tabbing points (white circles) to carry the current from one cell to the next in a module.

Corrosion can affect the metal grid in variety of ways. For simplicity, we will discuss three specific type of grid corrosion. More complex corrosion pattern can be viewed as a superposition of these "elementary" processes. First, *grid finger thinning* shown in Fig. 1b causes the current to travel through a finger with reduced cross section (i.e. higher resistance). Second, *grid delamination* shown in Fig. 1c prevents current pick-up by the finger, so that the carriers must travel further through the semiconductor laterally to reach the un-delaminated section of the metal line. Third, *busbar solder bond failure* shown in Fig. 1d eliminates the tabbing points and current must travel through a longer, more resistive path to reach the next module. The key insight in this paper is that *these corrosion processes affect the module J-V characteristics differently*, and therefore the electrical J-V signatures can be used to infer the types of corrosion processes within the module. We will explain how variation of these markers may be misinterpreted as signatures of different degradation mechanisms.

## III. Simulation Framework

We used a commercial cell simulator called Griddler[TM] [14] to calculate the J-V curves of cells with variously corroded/delaminated finger and busbars. Griddler needs as an input a map of the busbar and fingers as a starting point of the simulation. Therefore, we first used AutoCAD software to draw the relevant (pristine or corroded) patterns of fingers, busbars and tabbing points on the front side of the wafer. The backside was presumed fully covered with an opaque metal. Griddler uses the front grid pattern as a guide to spatially resolve the cell into small segments. Each spatial segment (i.e. node) is represented by a double diode five-parameter compact model (see Fig. 2a). The five parameter equation is given by:

$$J = J_{ph}(x,y) - J_{01}(x,y)(e^{\frac{qV(x,y)+J R_s(x,y)}{k_B T}} - 1) - J_{02}(x,y)(e^{\frac{qV(x,y)+J R_s(x,y)}{k_B T}} - 1) - \frac{V(x,y) + J R_s(x,y)}{R_{sh}(x,y)} \quad (1)$$

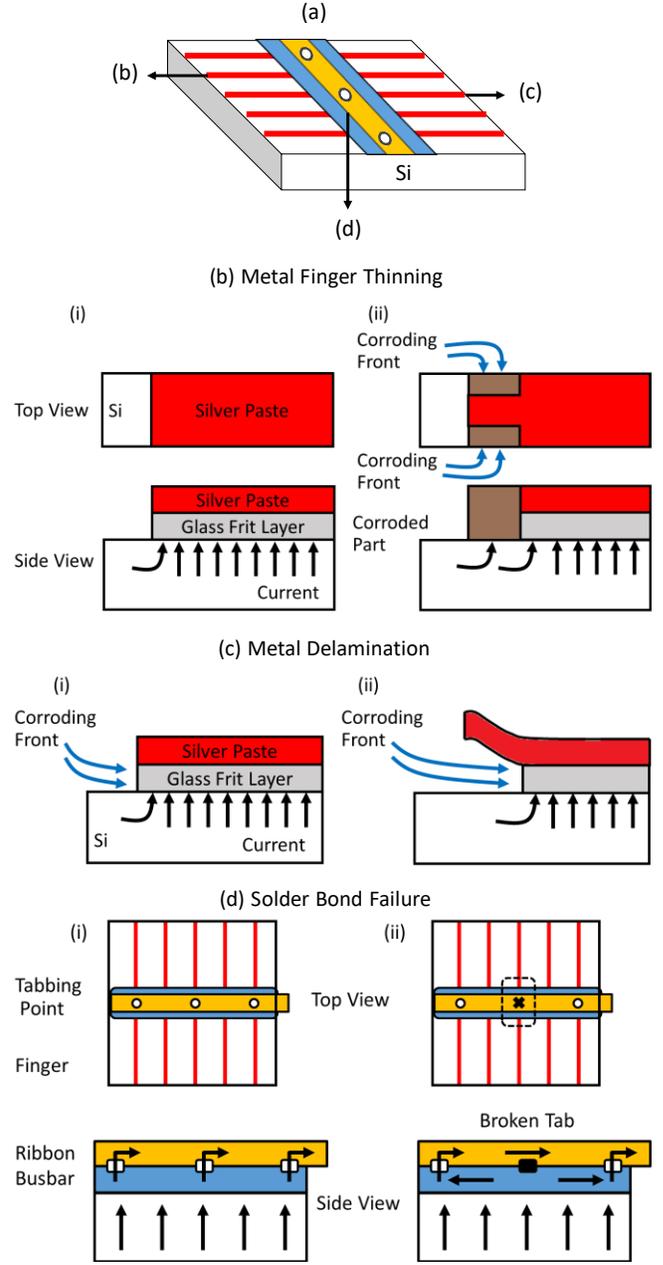

Fig. 1. (a) A schematic of a section of a solar cell and locations where three types of degradation may occur. Simple schematics of how corrosion changes the metal contacts and current path in the fingers, busbars, ribbon, and Si wafer when (b) metal thinning (c) metal delamination and (d) solder bond failure occur. (i) Before degradation, (ii) After degradation. (The glass frit allows electrical contact to the silicon surface after deposition of the anti-reflective coating during contact firing.[15])

The five parameters are: (a) $J_{ph}(x,y)$, the photocurrent density generated due to local illumination; (b) $J_{01}(x,y)$ is the diode recombination current with an ideality factor of one; (c) $J_{02}(x,y)$ is recombination current in the depletion region of the diode with an ideality factor of two; (d) $R_{sh}(x,y)$ is the local shunt resistant; and (e) $R_s(x,y)$, the series resistance. The baseline values that we used for different variables in the Griddler simulator are summarized in Table I. The variables are chosen for a typical c-Si cell with efficiency of 18.8%.

Once the *local* five-parameter model is specified, Griddler connects the nodes in a two-dimensional grid (see Fig. 2b) with appropriate front/back contact resistance and then solves the Kirchhoff's equations related to the network with appropriate boundary conditions (e.g., voltage and current specified at the end of the busbars). A self-consistent solution of the Kirchhoff's law allows one to obtain spatially resolved map of voltage and current distributions, $V(x,y)$ and $J(x,y)$. This map allows us to interpret cell J-V characteristics in terms of local processes.

For example, Fig. 2c shows the calculated voltage distribution for a pristine cell with two vertical busbars and a series of 82 horizontal fingers. The current is extracted at the bottom edge of the busbars (marked A and B), therefore these two points at the cell output (or module voltage, $V_c$) has the smallest voltage ($V_{min}$=0.555V, black). The busbars are essentially at the same potential (vertical black lines), because the resistances of the busbar is small. The local voltage differential

$$\Delta V(x,y) \equiv V(x,y) - V_{min} = V(x,y) - V_c < \Delta V_{max}$$

depends on the distance from the initial charge collection point on the finger (e.g. C, D, and E) to the final extraction point (i.e. A and B) at the busbar. At the farthest point from the busbar (C, D, E), the voltage is the highest ($V_{max}$ =0.596V, white), with the corresponding maximum voltage differential ($\Delta V_{max} = V_{max} - V_{min} \sim 40$ mV). Thus, for a pristine module, $\Delta V_{max} \sim 1.5 \frac{k_B T}{q} > \Delta V(x,y)$, therefore charge collection from various points within the cell is essentially uniform, because exponential terms on the right hand side of Eq. 1 are essentially identical, i.e. $V(x,y) \equiv V_{min} + \Delta V(x,y) \sim V_{min}$. This conclusion does not hold once the fingers/busbars begin to corrode, with dramatic implications for cell performance, and J-V characteristics.

TABLE I
BASELINE SIMULATION PARAMETERS FOR CELLS IN GRIDDLER

| Property | Value |
| --- | --- |
| Finger sheet resistance | $3\ m\Omega/sq$ |
| Busbar sheet resistance | $3\ m\Omega/sq$ |
| Finger contact resistance | $0\ m\Omega/sq$ |
| Layer sheet resistance | $80\ \Omega/sq$ |
| Wafer internal series resistance | $0\ m\Omega.cm^2$ |
| Internal shunt conductance | $0\ 1/(\Omega.cm^2)$ |
| Ribbon width | $1\ mm$ |
| Ribbon sheet resistance | $0.1\ m\Omega/sq$ |
| Contact point resistance | $0\ m\Omega$ |
| Number of tabbing points on ribbon | 15 |
| 1-sun $J_{ph}$, non-shaded area | $39.6\ mA/cm^2$ |
| $J_{01}$, passivated area | $200\ fA/cm^2$ |
| $J_{01}$, metal contact | $600\ fA/cm^2$ |
| $J_{02}$, passivated area | $10\ nA/cm^2$ |
| $J_{02}$, metal contact | $50\ nA/cm^2$ |
| Finger pitch | $1.9\ mm$ |
| Finger width | $60\ \mu m$ |
| Number of fingers | 82 |
| Busbar width | $1.5\ mm$ |
| Number of busbars | 2 |
| Cell area | 156x156 $mm^2$ |
| Front illumination | 1 Suns |

## IV. EFFECT OF FINGER CORROSION AND SOLDER BOND FAILURE

When a module is installed in field, it is exposed to environmental humidity. Moisture first penetrates the module through cracked sealant, power plugs, (fractured) glass, and/or cracked backsheet. Subsequently, moisture diffuses through the encapsulant to eventually reach the solar cell. Moisture may corrode the fingers/busbar of a solar cell in two ways. First, in the *dark-corrosion* mode, moisture initially reacts with the encapsulant EVA to produce acetic acid [16]. The acid dissolves parts of metal in contact with silicon and the "delamination" increases the series resistance [17]. In the *light-corrosion* mode during normal daytime operation, moisture is directly hydrolyzed at the metal contacts and the $OH^-$ molecule reacts with the metal to produce metal hydro-oxides ($MOH^-$). The hydro-oxides is eventually neutralized by PID-related $Na^+$ ions from the front glass [18]. This dissolution of metal or the delamination of the electrode due to build-up of $H_2$ gas (hydrolysis product) degrade current collection. The actual reactions are complex and still subject to intense study [18]–[21].

In this paper, we will not focus on the kinetics of corrosion, but rather analyze the J-V signature of CDS once it has occurred. As discussed, corrosion can thin the grid finger by dissolving away the metal from the cell edges located close to the module edge (Fig. 3) or cause delamination of the contacts along the length of the finger (Fig. 5). We will explain the effects of each corrosion pattern in the following subsections. In addition, the busbar solder bond failure that occurs due to thermal expansion and contraction will be discussed with reference to Fig. 7.

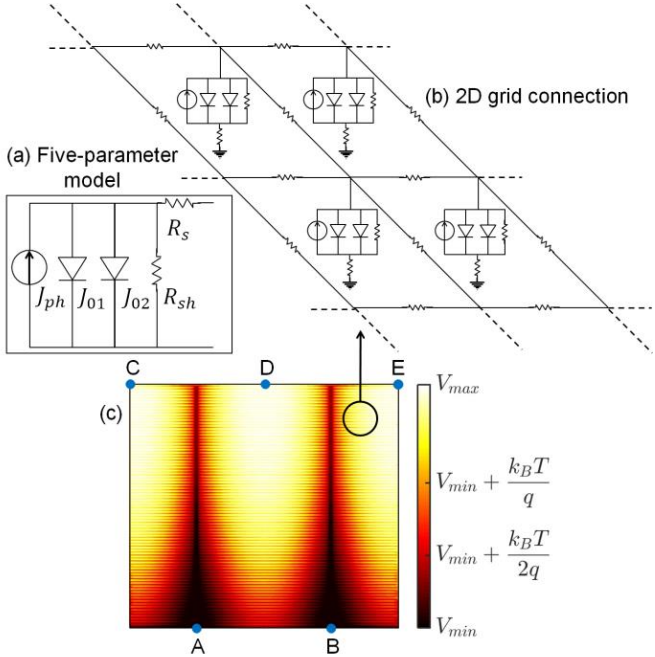

Fig. 2. (a) A schematic of the Five-parameter model. (b) Two-dimensional grid connection of the five-parameter model in Griddler. The magnitudes of the circuit elements depend on the location within the cell. (c) Voltage map distribution of a pristine cell. The difference between $V_{min}$ and $V_{max}$ is in order of $k_B T/q$, allowing highly efficient current collection from any location within the cell.

*A. Metal Finger Thinning*

Thinning of the metal fingers occurs when the corrosion attacks the contacts from the edges. For simplicity, we assumed the thinning happens uniformly for a length of 3.0 cm (see Fig. 3a). This pattern may arise because corrosion is slower than moisture diffusion, although exact details of the corrosion pattern will not affect the key insights presented in this paper. The simulation set-up only affects the finger resistance. Therefore one expects that the changes of the J-V characteristics would be attributed to increase in effective series resistance, $R_s$.

Interestingly, the thinning of the fingers **does not** affect J-V curve until five-sixth of the initial width (60 $\mu m$) is dissolved (See Fig. 4). The fingers are overdesigned to ensure high yield, therefore despite initial corrosion reducing the width of the finger, the voltage drop along the finger $(\Delta V(x,y))$ is relatively small and the voltage redistribution (Fig. 3b) and the cell performance (Fig. 4a) appears indistinguishable compared to pristine cells.

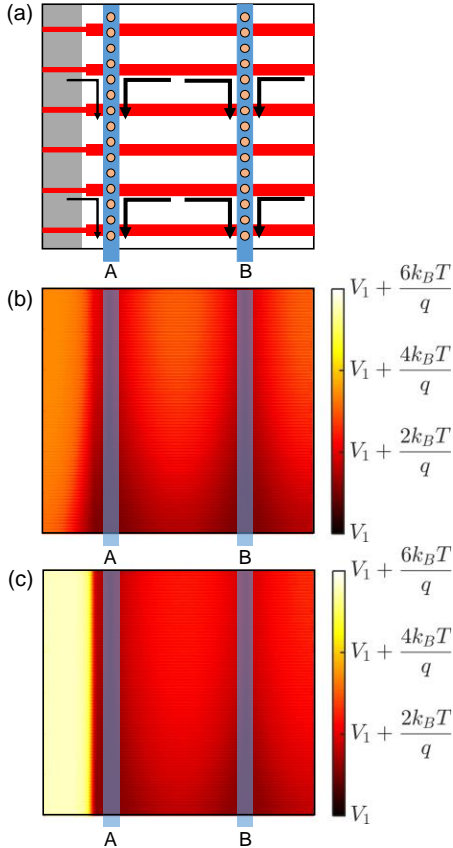

Fig. 3. (a) A schematic diagram of grid configuration associated with finger thinning. (b) and (c) show corresponding voltage map at $V_{MP}$ where 40.0 $\mu m$ and 59.5 $\mu m$ out of 60.0 $\mu m$ of the finder width have been corroded. Arrows show a general current collection path. Thinner arrows indicate less current is being collected due to finger thinning. Shaded area in (a) corresponds to the brighter part in (b) and (c) that have higher voltage drop. A large portion of current in shaded area is being sunk in the strongly turned on diodes and does not contribute to total collected current.

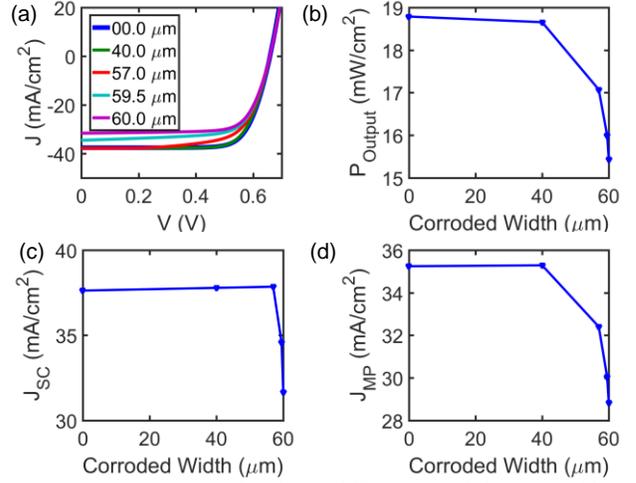

Fig. 4. J-V curves associated with five different corroded finger widths. The output power follows the $J_{MP}$ trend. The slight $J_{SC}$ increase is due to partial removal of the contacts and reduced shading of contacts. Significant output power reduction starts after 50 um corrosion of the finger width.

The voltage distribution (Fig. 3b) and the J-V characteristics (Fig. 4a) begin to change significantly when the finger width is reduced to 10-20 $\mu m$). Now the finger current travels laterally through the high resistance corroded section (or the semiconductor underneath) until it reaches the healthy section of the finger. The voltage over corroded section in Fig. 3b approaches $V_{max} = 0.610V$ (white) or $\Delta V_{max} \sim 60$ mV $> 2\frac{k_BT}{q}$. As $V_c(= V_{min})$ is increased, the distributed diodes in the corroded section are turned on even more strongly (due to their exponential dependence of $\Delta V(x,y)$, as in Eq. 1). The J-V characteristic in Fig. 4a gives the appearance of a weakly "shunted" cell. The output power is reduced as diodes now dissipate the local photocurrent instead of allowing them to be collected by the busbar.

Finally, when most of the finger is corroded (e.g., 59.5 $\mu m$ out of 60 $\mu m$), $\Delta V_{max} = 110$ mV (Fig. 3c, white section on the left) ensure that strong diode turn-on dissipates most of the local photo-current within the corroded section. The short-circuit current is reduced in a manner similar to yellowing or partial vertical shading. The situation is worse, because the hot spots formed will accelerate corrosion/delamination.

To summarize, initial corrosion may not affect the J-V characteristics at all. Intermediate corrosion (40-55 $\mu m$) is reflected as a **fake shunt resistance** (indicated by the slope at V=0) at the terminal J-V characteristics. At final stages of corrosion, the J-V characteristics resemble partial shading and local yellowing. Thus, finger corrosion is susceptible to mischaracterization at the J-V level. This misinterpretation is especially dangerous when one wishes to differentiate Potential Induced Degradation (PID) (that affects the shunt resistance) from finger corrosion solely based on the terminal J-V characteristics.

*B. Metal Delamination (Grid Dissolution Under Dark)*

The dark and light corrosion can delaminate (or fully corrode) the fingers and busbars from the semiconductor underneath. For simplicity, let us assume that moisture diffuses uniformly from the side close to the module edge. Therefore, the fingers are fully delaminated the same length (see Fig. 5a), which is equivalent to the worst scenario in metal finger thinning. Fully delaminated fingers cause the current to travel laterally towards the healthy parts of the finger through the semiconductor and be collected by the remaining fraction of the finger. Since the semiconductor resistance is relatively higher, $\Delta V(x,y) \gg \frac{k_B T}{q}$ is significant in the white delaminated section of the cell as shown in Fig. 5b. The turned-on diodes shunt the current, almost none of the photo generated carriers are collected from this region, the short-circuit current is reduced (Fig. 6c) and the power drops linearly with delaminated area. This effect is similar to yellowing or shading that also reduce the $J_{SC}$ without any change in FF, again suggesting possibility of mischaracterization. Visual inspection or IR imaging will differentiate the degradation modes: yellowing does not produce local hot spots, but grid delamination do.

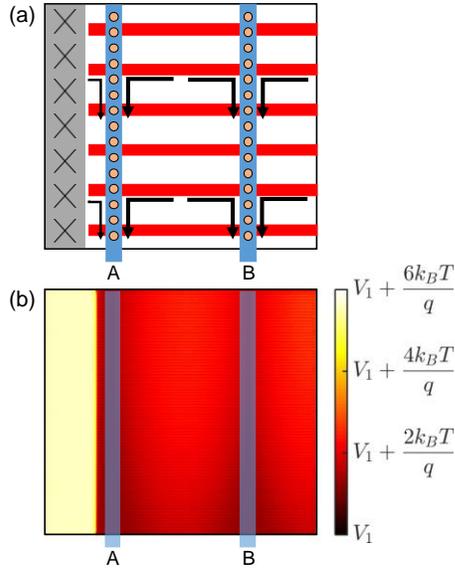

Fig. 3. (a) Finger delamination schematic and (b) voltage map at $V_{MP}$ where $3\ cm$ out of $15.6\ cm$ finger length has been delaminated. Arrows show a general current collection path. Thinner arrows show less current is being collected due to finger delamination. Shaded area in (a), corresponding to brighter part in (b), does not contribute to total collected current. All of the current in this part marked by "x" is being sunk in the strongly turned on diodes and turns to heat.

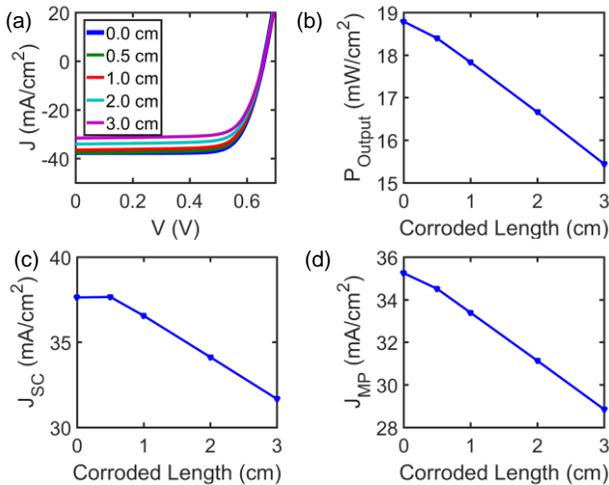

Fig. 6. J-V curves for five different corroded lengths. The output power follows the $J_{MP}$ trend. The voltage drop is not significant when 0.5 cm of length is corroded, therefore, the $J_{SC}$ does not drop.

In closing this section, let us reiterate that the series resistance has remained essentially unchanged even in this extreme case of finger corrosion. *Does corrosion ever lead to series resistance increase, as presumed in the traditional literature?* Yes, it does but only for solder bond failure – the topic of next section.

### C. Solder Bond Failure

Thermal expansion/contraction during thermal cycling or hourly/daily/seasonal temperature variation can cause solder bond failure. To model the solder bond failure in Griddler, we sequentially remove the tabbing points between the ribbon and the busbar. The current is now forced to take alternate paths in higher resistance busbars, leading to a fundamentally altered voltage distribution pattern in the cell.

Pristine cell voltage map $V(x,y)$ is shown Fig. 2c, with ribbons and tabbing points intact. When a tabbing point fails, the current reroutes through the busbar and reaches the adjacent tabbing point (see Fig. 1d). If the failed tabs are consecutive and close to current collection plugs (bottom of the cell at points A and B), this rerouted path leads to large resistive drop, with dramatic increase in $\Delta V(x,y)$, as shown in Fig. 7b (7 bottom tabbing point removed). The arrows indicated the rerouting of current flow. As a result, the photo-generated current at the bottom edge of the cells (white region) may not be as efficiently collected. Interestingly, as the first tabbing points starts at the "middle" of the cell, the voltage is redistributed leading to more efficient charge collection from the remainder of the cell, as shown in Fig. 7b.

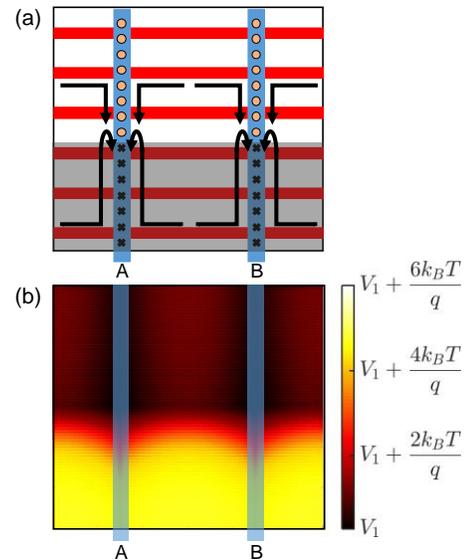

Fig. 7. (a) Solder bond failure schematic and (b) voltage map at $V_{MP}$ where 7 bonds (tabbing points) out of 15 have been broken. Arrows show a general current collection path from fingers to busbars and then ribbons. The portion of current in shaded area (a) goes through a longer and more resistive path in the busbar to reach the first healthy tabbing point. Due to the metallic resistivity of the path that current goes through the voltage drop is not high enough to turn the diodes on and sink current significantly at low bias.

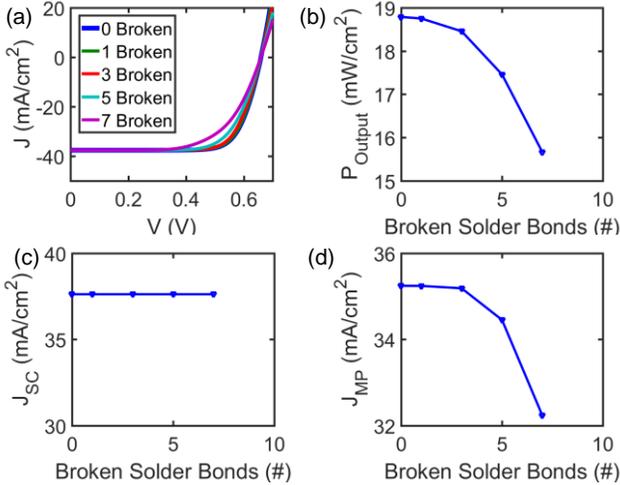

Fig. 8. J-V curves as a function of the number of broken solder bonds. Since the voltage drop is not significant at $J_{SC}$ there is no change in $J_{SC}$ even when almost half of the bonds are broken. However, the $J_{MP}$ drops as the voltage drop over the broken bonds increases.

Fig. 8a shows that solder bond failure manifests a distinct signature of series resistance both in FF and slope of the J-V curve close to $V_{oc}$. This is because any corrosion at the tabbing point and/or ribbons affects not the local current, but the integrated current of the cell. In addition, close to $V_{MP}$ when the voltage drop is high enough for the distributed diodes to turn on, there is a significant loss of current collection. However, the drop is not high enough at $J_{SC}$, thus no reduction in $J_{SC}$.

## V. Cell vs. Module J-V Characteristics

Degradation mechanisms such as corrosion reduce output power at the cell level, but how does corrosion affect the module performance that contains a combination of corroded and healthy cells? After all, qualification or field tests are done on encapsulated module where the J-V characteristic of individual cells are not available. Since only a few cells of a module may be affected by CDS, it is important to know how the cell level degradation translates to module-level J-V characteristics. Many field inspections report degradation happens at the edges of the module [22]–[24], therefore, Fig. 9a shows $N$ degraded cells along the edges of the module, with $M$ healthy cells at the interior of the module. The exact location of the degraded module in this series connected system is unimportant, so long $M$ and $N$ are specified. As shown in Fig. 9b-d, we find that the features of cell J-V characteristics (following CDS degradation) is preserved in module J-V characteristics. This occurs despite the complexity of the voltage and current redistributions among the degraded and pristine cells. For example, Fig. 9b shows that finger thinning at the cell level appears as a shunt resistance even at the module level. Similarly, finger delamination at the cell level translates to suppression of the module short-circuit current; see Fig. 9c. Finally, busbar corrosion leads to an increase in the series resistance, as in Fig. 9d. Note that the presence of healthy cells reduces the magnitude of the series resistance seen at corroded cell J-V characteristics. When the effects are present simultaneously, (i.e. some cells are delaminated, while others have lost the tabbing contact) the module characteristics is defined by a convolution of the elemental features.

## VI. Discussion

Since CDS manifest variously as $R_{sh}$ decrease (finger thinning), $J_{sc}$ loss (delamination), and/or $R_s$ increase (solder bond failure), it is important to distinguish the signature of CDS from other degradation mechanisms, such as PID ($R_{sh}$ decrease) and/or shadowing or yellowing ($J_{SC}$ loss). First, it is clear that $R_s$-increase can always be positively correlated to solder-bold failure. The other effects may be differentiated by the following electrical/optical characterization methods.

### A. Reverse bias characteristics

Finger corrosion related "diode shunts" saturates at sufficiently high reverse bias, while PID related increase in "real" shunt current increases with reverse bias. Thus, reserve bias characteristics will differentiate between the two degradations. In addition, PID on its own does not change the short-circuit current; however, finger corrosion may affect $J_{sc}$ depending on its severity (as shown in Fig. 4). A Griddler simulation of the under illumination J-V shown in Fig. 10 demonstrates the feasibility of this approach. In practice, the approach may require removal of the protection diodes from a module and carefully ensure that the cells do not go into reverse breakdown.

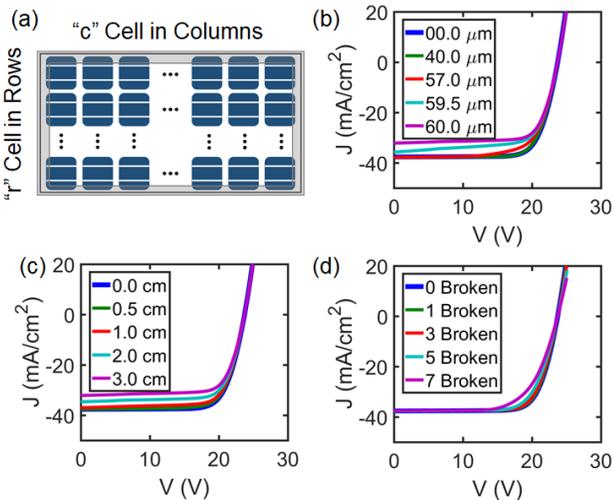

Fig. 9. (a) A schematic of the degraded cells within a module. Here, $N$ is the number of degraded cells $N = (r + c - 2) \times 2$, whereas $M$ is the number of healthy cells $M = r \times c - N$ ($M = 10, N = 26$). We assume only the edge cells in the gray area degraded due to corrosion. Module level J-V curves synthetized using cell level results of Griddler J-V curves for different degradations involving: (a) metal finger thinning, (b) finger delamination, and (c) solder bond failure.

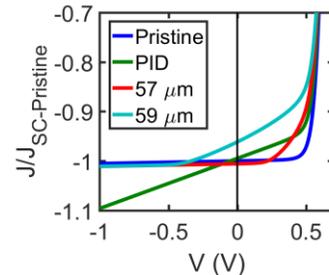

Fig. 10. J-V curves compare PID vs. various degrees of finger thinning of 57 $\mu m$ and 59 $\mu m$ out of 60 $\mu m$ width. Same area of the cell has been affected by the two degradation mechanisms.

*B. Optical measurement to distinguish between yellowing and delamination-induced loss in photo-current*

Delamination-induced $J_{SC}$-loss leads to local dissipation of photocurrent and corresponding localized increase in self-heating and cell temperature. Yellowing prevents photons from reaching the cell, and thus the loss of photo-generated current is not related to local hot-spot formation. Therefore, an IR image should differentiate between these two degradation modes. Another way to distinguish between the two is visual inspection. Yellowing generally affects most of the area of the module/cell; however, delamination normally occurs close to the edges of the module/cell. Finally, one may analyze the J-V curves at reverse bias. As shown in Fig. 11 yellowing reduces the short circuit current, but does not affect the slope of the curve at reverse bias. Delamination on the other hand, not also reduces the short-circuit current, but also feature a shunt like signature at reverse bias associated with gradual turn-off of the diodes far from the current collection points.

*C. Combination of degradation mechanisms*

The discussed methods can distinguish the mechanisms when only one is influencing the cell/module. If two or more mechanisms affect the J-V curve simultaneously (e.g. yellowing and PID) the features of J-V characteristics may be mischaracterized as being due to delamination. Therefore, it may not be possible to distinguish between the mechanisms only relying on electrical characterization. Other methods such as visual inspection and/or IR imaging will be helpful. Yellowing uniformly affects the module and is easy to spot by visual inspection [24]. IR imaging will spot the shunted areas due to higher recombination in the PID affected parts. Thus, it is possible to differentiate between the mechanisms based on electro-optical multi-probe characterization methods.

## VII. SUMMARY AND CONCLUSIONS

In this paper, we have used Griddler, a solar cell and module simulator, to investigate the effects of CDS on performance of solar cells. We find that: (a) only solder bond failure is directly and positively correlated to increase in the series resistance, as characterized by the derivative close the open-circuit condition; (b) finger thinning does not affect the performance of cell significantly until the finger width reduces to less than 10 $\mu m$. The electrical signature of finger thinning is a steeper slope in lower voltages, which may be mischaracterized by shunt resistance. Reverse bias J-V characteristics may be used to distinguish between diode shunt due to corrosion, and actual shunt due to PID, for example. Finally, finger delamination reduces the performance by sinking the photo-generated current locally and reduces $J_{SC}$. An optical image, IR image of hot-spot formation, and a slope of J-V curve at reverse bias help differentiate between yellowing and finger delamination. A deep and nuanced understanding of complex correlation of a degradation mode and its electrical signatures (reflected in the J-V) characteristics is essential for interpreting the qualification tests and fields results and improve the next generation of solar modules manufactured for a specific weather zone.

Acknowledgment

This work was supported by the National Science Foundation under Grant No. 1724728.

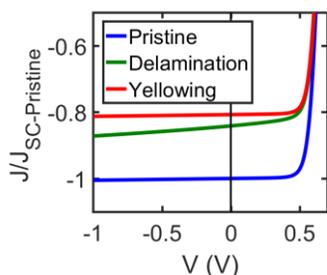

Fig. 11. J-V curves compare yellowing and delamination, where 3 $cm$ out of 15.6 $cm$ long finger has been delaminated. Yellowing reduces the short circuit current; adding delamination results in a shunt like signature.